\documentclass[preprint,amsmath,titlepage,amssymb,english,aps,showkeys]{revtex4}
\usepackage[utf8]{inputenc}
\usepackage[english]{babel}
\usepackage{graphicx}
\usepackage{latexsym,amsmath,amsthm,amssymb,amsfonts}
\usepackage{makeidx}   
\usepackage{fancyref,fancyhdr,pifont}
\usepackage[T1]{fontenc}
\usepackage{subfigure}
\usepackage{graphics}
\usepackage{amsmath}
\usepackage{dcolumn}
\usepackage{amssymb}
\usepackage{bm}
\usepackage{cancel}
\usepackage{xcolor}
\usepackage{hyperref}

\def\sech{\mathrm{Sech}}
\def\csch{\mathrm{Csch}}
\def\coth{\mathrm{Coth}}
\def\sinh{\mathrm{Sinh}}
\def\cosh{\mathrm{Cosh}}
\usepackage{textgreek}
\usepackage{comment}
\usepackage{upgreek}
\def\pa{\partial}
\begin{document}
	\title{Collective coordinates for the hybrid model}
	
	\author{C. F. S. Pereira}
	\email{carlos.f.pereira@edu.ufes.br}
	\email[,]{etevaldocostaf@gmail.com}
	\email[,]{tadeutassis@gmail.com}
	\affiliation{Departamento de Física e Química, Universidade Federal do Espírito Santo, Av. Fernando Ferrari, 514, Goiabeiras, 29060-900, Vit\'oria, ES, Brazil.}

\author{Etevaldo dos Santos Costa Filho}
	\affiliation{Departamento de Matemática da Universidade de Aveiro and
		Centre for Research and Development in Mathematics and Applications (CIDMA),
		Campus de Santiago, 3810-183 Aveiro, Portugal}

	\author{T. Tassis}
	\affiliation{Centro de Ciências Naturais e Humanas, Universidade Federal do ABC, Av. dos Estados, 09210-580, Santo André, São Paulo, Brazil}

	\begin{abstract}
	 
In the present work, we carry out the study of scattering solitons for the \textit{anti-kink/kink} and \textit{kink/anti-kink} configurations. Furthermore, we can observe the same effects as those described by D. Bazeia et al. \cite{hibrid}. We apply the collective coordinate approximation method to describe both scattering configurations and verify that just as happens in the polynomial models $\phi^4$ and $\phi^6$ \cite{weigel, taky, belova, tadau, camp}, the method has its limitations regarding the initial scattering speeds. In such a way that, for certain initial speeds, the solution of collective coordinates agrees with the fullsimulation, and for other speeds, there is a discrepancy in the solutions obtained by these two methods. We also noticed that, considering the hybrid model, the \textbf{null-vector} problem persists for both configurations, and when trying to fix it, a singularity is created in moduli-space as well as in $\phi^4$ \cite{manton1}.
	\end{abstract}
	
	\keywords{Solitons scattering, linear fluctuations, internal modes, and collective coordinates.}
	
	\maketitle
	
	\section{Introduction}\label{sec1}

	In the last decades, the interest in the study of topological solitons has grown considerably. This interest is due to the wide applicability of such solutions in several domains of physics, which can be found, for instance, in high energy physics \cite{manton, raja}, condensed matter \cite{matter}, cosmology, gravitation \cite{vilen,vacha,weiberg} and even in the study of optical fiber \cite{fibra} and in biological systems \cite{witt,system}. Thus, the so-called Derrick theorem \cite{derrick} guarantees that the space dimension restricts the non-existence of static and topological solutions, being possible in only $(1+1)$ dimensions. However, there are several ways to get around this theorem and obtain topological solutions for higher dimensions. As is the case for vortex  \cite{vortexp} and Q-ball  \cite{q1,q2,q3} solutions in which gauge coupling is introduced and in skyrmion models \cite{ski1,ski2} in which terms derived from higher orders are entered.

As its fundamental principle, the collective coordinate method transforms a field theory problem into one of classical mechanics. For that, all its spatial degrees of freedom must be integrated throughout the space so that the new Lagrangian functional will only depend upon the dynamical parameters, that is to say, the dynamics variables of the system and its derivatives. In this perspective, the technique has been extensively explored in the study of scattering of solitons in $(1+1)$ dimensions, with greater emphasis on the $\phi^4$ model. In recent years, several authors have been dedicated to trying to describe the occurrence of energy transfer from translational to vibrational modes and how this redistribution occurs so that the solitons go directic to the asymptotic \cite{camp,belova,good,moshi,samuel,tadau}. Concomitantly, it rekindled interest in the study of this model since some authors identified \cite{taky,weigel,goodnovo,carlosp} a crucial error in some of the functions necessary for the lagrangian functional to be written explicitly. The procedure was also applied to the study of the classical 2-\textit{kinks} scattering in the sine-Gordon \citep{cliff}, and modified sine-Gordon  \cite{helen} models, as well as, for the nonlinear Schrödinger model \cite{gabi} in which the collective coordinate was introduced as a temporal phase.

As for the model $\phi^6$ in $(1+1)$ dimensions, it was already known through numerical experiments by \cite{dorey} that the scattering between \textit{anti-kink/kink} had states of resonance and along with the effect the existence of collective vibrational modes. Thus, in \cite{gani} the authors sought to build the collective coordinates numerically for all scattering scenarios. They realized that the \textit{anti-kink/kink} case in which it should be possible to have some information about the resonance windows. However, for the \textit{ansatz} used, it was not possible to be verified.

In the present work, we will apply the technique of collective coordinates to study a hybrid model that was presented by D. Bazeia et al. in \cite{hibrid}. This particular model has an intersection of the polynomial model $\phi^4$ with  $\phi^6$, and for specific velocities regimes, new interesting structures can be explored. The idea of doing the study using the method of collective coordinates aims to seek a qualitative and quantitative treatment to understand the effects that appear in the formation of the resonance windows. In short, we want to investigate how the energy balance occurs between its degrees of freedom (translational and vibrational modes).

The paper is organized as follows: we start the work in Section 2 by presenting the background and the model's basis. Next, in Section 3, we apply the technique of collective coordinates for scattering \textit {anti-kink/kink}. In Section 4, we construct the collective coordinates for scattering \textit{kink/anti-kink}; in section 5, we investigate the \textbf{null-vecto}r problem and finally, in Section 6, the conclusions are presented. 
	
	\section{The model}\label{sec2}
	
	As said before, the hybrid model, as proposed in \cite{hibrid}, is given as a certain combination of the models $\phi^4$ and $\phi^6$. The action for such model is given by
	
	\begin{eqnarray} \label{1}
	S_{Hybrid}=\int{d^2x}\left(\frac{1}{2}\partial_{\mu}\phi\partial^{\mu}\phi-V(\phi)\right),
	\end{eqnarray} 
	\begin{eqnarray}
	V(\phi)=\frac{\lambda\phi^2}{4}\left(\frac{m}{\sqrt{\lambda}}-|{\phi}|\right)^2.
	\end{eqnarray}
	We deal with a real Lorentz scalar field, $\phi$, in $1+1$ dimensional Minkowski space-time. Here, the parameters $m$ and $\lambda$ are chosen to be real. Notice that the above potential possesses three vacuum states given by $\phi=0$ and $\phi=\pm \dfrac{m}{\sqrt{\lambda}}=\pm\eta$, and each of these vacua possesses multiplicity two. Moreover, these three vacua connect two distinct topological sectors. See figure below 

	\begin{figure}[h]
		\centering
		{			\includegraphics[scale=0.6]{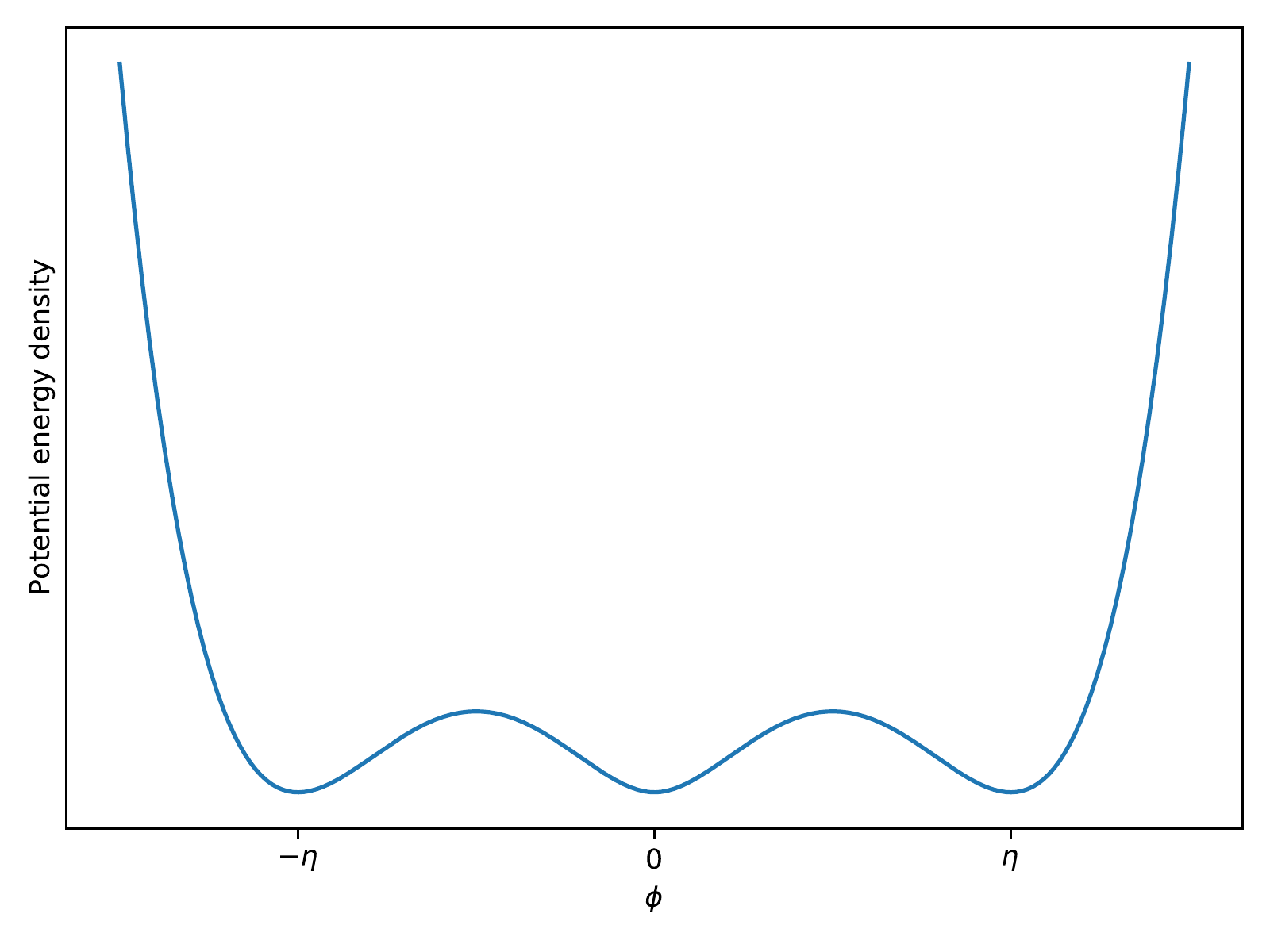}}
		\caption{We plot here the potential $V(\phi)$.}
		\label{potentialform}.
	\end{figure}
	
	The equation of motion of the field is given by
	
	\begin{equation}
	\pa_\mu\pa^\mu\phi+\dfrac{\lambda\phi}{2}\left(\dfrac{m^2}{\lambda}-\dfrac{3m|\phi|}{\sqrt{\lambda}}+2\phi^2\right)=0,
	\end{equation}
	
	Without loss of generality, we will follow the convention on the literature and set from now on $m^2=\lambda=2$ in order to properly do the numerical calculations. The \textit{kink/anti-kink} static configurations for each of these two sections, connecting the vacua, are giving by the equations:
	
	\begin{eqnarray}\label{2}
	\phi_{K}^{\left(0,1\right)}&=& \frac{1}{2}\left[1+\tanh\left(\frac{x}{2}\right)\right]  \qquad \qquad \phi_{\bar{K}}^{\left(1,0\right)}= \frac{1}{2}\left[1-\tanh\left(\frac{x}{2}\right)\right]\\
	\phi_{K}^{\left(-1,0\right)}&=& \frac{1}{2}\left[-1+\tanh\left(\frac{x}{2}\right)\right] \qquad \qquad \phi_{\bar{K}}^{\left(0,-1\right)}= -\frac{1}{2}\left[1+\tanh\left(\frac{x}{2}\right)\right].
	\end{eqnarray} 
	
\begin{figure}[htb]
		\centering
		\mbox{%
			\subfigure[]
			{\label{kinks}%
				\includegraphics[scale=0.47]{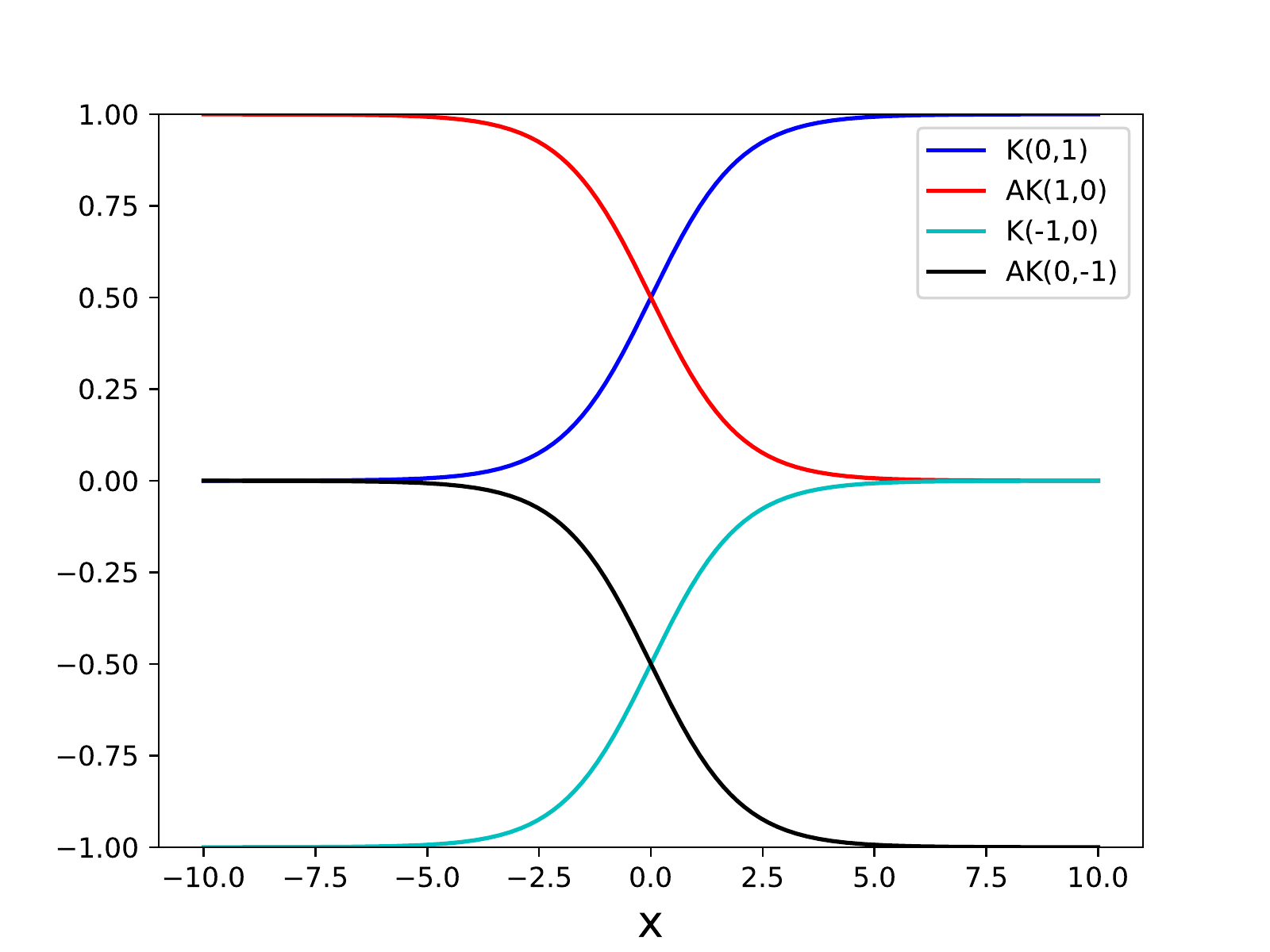}}\qquad
			\subfigure[]
			{\label{self}%
				\includegraphics[scale=0.47]{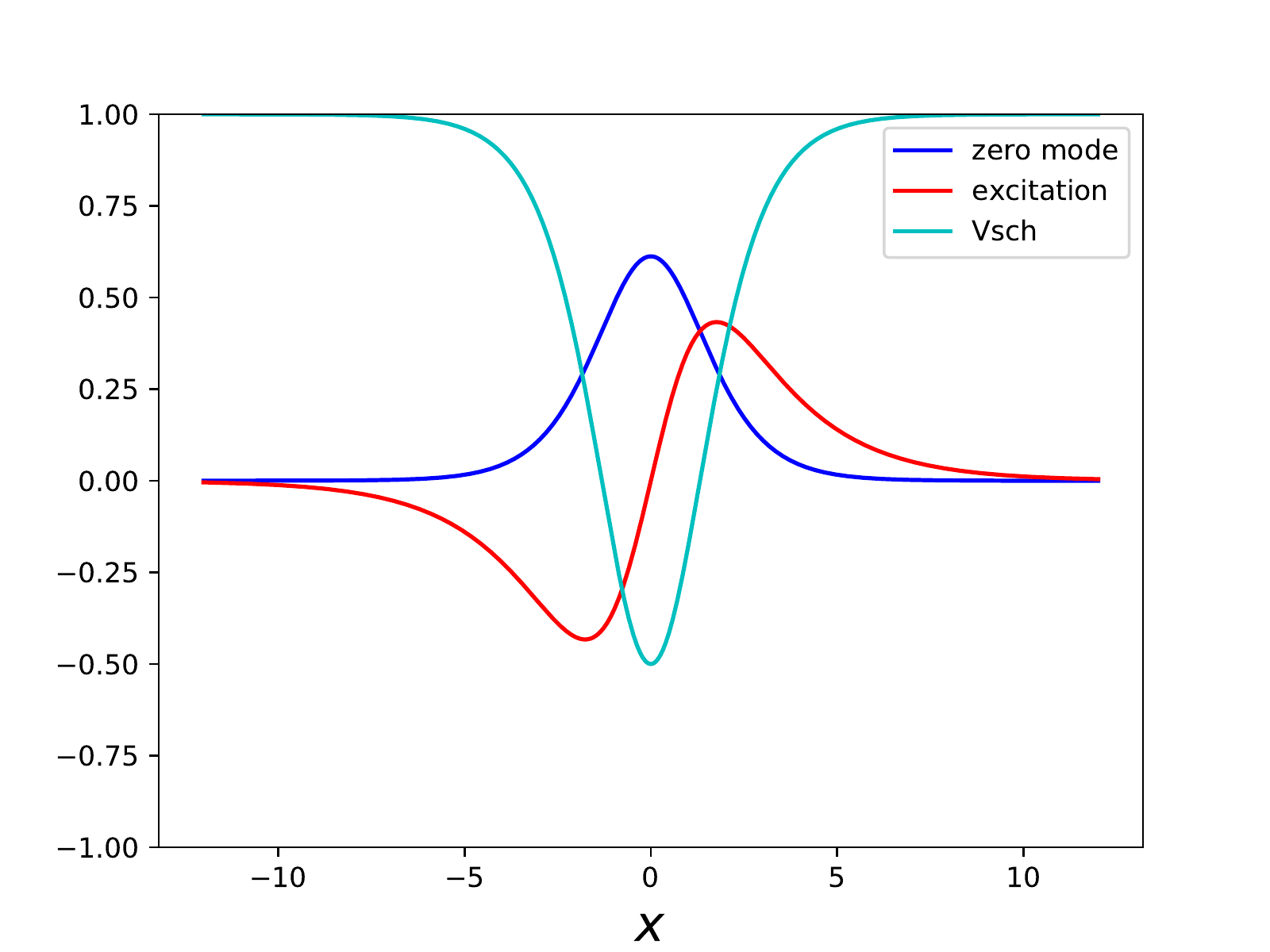}}
		}
		\caption{(a) Field configuration for the \textit{kink} and \textit{anti-kink} b) we have the graphic referring to the effective potential arising from linear pertubations and the curves for the translational and vibrational mode.}
		\label{fig1}.
	\end{figure}

	Notice that such configuration is slightly similiar to the $\phi^6$ model see \ref{kinks}. Considering linear perturbations around these static vacuum solutions, $\phi\left(x,t\right)=\phi_{K,\bar{K}}\left(x\right)+ \chi\left(x\right)\cos\left(\omega{t}\right)$, we find a Schrödinger-like equation for the spatial perturbation component  see \ref{potentialform} possessing the function $V_{Sch}$ as potential,
	
	\begin{eqnarray}\label{3}
	V_{Sch}= \left(-\frac{1}{2}+\frac{3}{2}\tanh^2\left(\frac{x}{2}\right)\right).
	\end{eqnarray} 
	
	Notice the existence of a map between the eigenfunctions of the present model with those obtained through the   Pöschl-Teller potential for the $\phi^4$ polynomial model. In this way, their eigenfrequencies are related as given by the expression $\omega_{Hibrid}^2=\frac{1}{4}\omega_{\lambda\phi^4}^2$. Therefore, the eigenfunctions associated with the translational and vibrational modes are giving, respectively, by the equations \eqref{4},\eqref{5}.

	\begin{eqnarray}\label{4}
	\omega_{0}^2=0  \qquad \qquad \longrightarrow \qquad \qquad   \chi_{0}= \sqrt{\frac{3}{8}} \sech^2\left(\frac{x}{2}\right), 
	\end{eqnarray}
	
	\begin{eqnarray}\label{5}
	\omega_{1}^2= \frac{3}{4} \qquad  \qquad \longrightarrow \qquad \qquad  \chi_{1}= \sqrt{\frac{3}{4}} \tanh\left(\frac{x}{2}\right)\sech\left(\frac{x}{2}\right). 
	\end{eqnarray}
	
	This effective potential, $V_{Sch}$, which arises from the perturbations,  is shown in figure \ref{self} together with the translational and vibrational modes.
	
	\section{Collective coordinates for the \textit{anti-kink/kink} scattering}\label{sec3}
	\subsection{Numerical Inputs}
	
	For all numeric simulations, we used the python library "python scipy.integrate". Some numerical values are settled as follows. For the spatial and temporal grid, it was sufficient to use $dx=0.05$ and  $dt=\frac{dx}{2}$. For both scattering, \textit{anti-kink/kink} and \textit{kink/anti-kink}, we fix the initial positions as $x=\pm{x_{0}}$, where  $x_{0}=15$. As regard the model constants, we set $m^2=\lambda=2$.
	
	\subsection{\textit{Anti-kink/kink} scattering}
	
It is equivalent to study the phenomenology of the scattering process of \textit{kinks} in the first topological sector or the second sector. In this way, we will only consider as reference the scattering process of \textit{kinks} that occurs in the first topological sector.

	Therefore, for the scattering \textit{anti-kink/kink}, we consider the following field configuration
	\begin{eqnarray}\label{7}
	\phi= \phi_{\bar{K}}^{\left(1,0\right)}\left(x+ a(t)\right)+\phi_{K}^{\left(0,1\right)}\left(x -a(t)\right) + \xi\left(t\right)\left[\chi_{\bar{K}}^{\left(1,0\right)}\left(x+a(t)\right)-\chi_{K}^{\left(0,1\right)}\left(x-a(t)\right)\right].
	\end{eqnarray}
	
	Here, the translation parameter $a(t)$ represents the relative distance between the center of mass of the \textit {kink} and the \textit {anti-kink}. By it turns, the parameter $\xi(t)$ is the amplitude of the vibrational mode. The vibrational mode of the  \textit{anti-kink/kink} solution is, respectively, $\chi_{\bar{K}}^{\left(1,0\right)}$ and $\chi_{K}^{\left(0,1\right)}$. 
	
	As it is well-known, the collective coordinates method is based on integrating the Lagrangian density in all spatial degrees of freedom of the field configuration \eqref{7}; and hence, to get a functional which only depends on the dynamic variables of the system and its derivatives. This is the procedure for describing moduli-space dynamics.

	In what follows, since we are studying perturbations of the vacuum, we will neglect second-order terms of the vibrational mode amplitude $(O\left(\xi^2\right))$.
	
	\begin{eqnarray}\label{8}
	L_{1}\left(a,\dot{a},\xi,\dot{\xi}\right)=\int_{-\infty}^{\infty}{dx}\left[\frac{1}{2}\left(\frac{\partial{\phi}}{\partial{t}}\right)^2-\frac{1}{2}\left(\frac{\partial{\phi}}{\partial{x}}\right)^2-\frac{1}{2}\phi^2\left(1-|\phi|\right)^2\right].
	\end{eqnarray}
	
	Considering the scattering, the effective Lagrangian is given by
	
	\begin{eqnarray}\label{9}
	L_{1}\left(a,\dot{a},\xi,\dot{\xi}\right)&=&\frac{1}{2}\left[\bar{M}+I_{1}\left(a\right)+\xi{J_{1}\left(a\right)}+\xi^2{K_{1}\left(a\right)}\right]\dot{a}^2 +\frac{1}{2}\left[{2+Q_{1}\left(a\right)}\right]\dot{\xi}^2 + \\\nonumber
	&+& \left(C_{1}\left(a\right)+\xi{N_{1}\left(a\right)}\right)\dot{a}\dot{\xi} -\left(V_{1}\left(a\right)-\xi{F_{1}\left(a\right)}+\xi^2{W_{1}\left(a\right)}\right),
	\end{eqnarray}Here $\bar{M}=2M_{0}$, where $M_{0}$ is the rest mass of the kink.
	
	The analytical expression of all functions above is given here,
	
	\begin{eqnarray}\label{10}
	\bar{M}=\frac{1}{3} \qquad \qquad \qquad W_{1}\left(a\to \infty\right)=\frac{3}{4}
	\end{eqnarray}
	\begingroup
	\small
	\begin{eqnarray}\label{11}
	I_{1}\left(a\right)&=& \csch^2\left(a\right)\left(-1+a\coth\left(a\right)\right)\\
	V_{1}\left(a\right)&=&-\frac{2}{3} + a + \frac{3}{\tanh\left(a\right)} -\frac{2(1+3a)}{\tanh^2\left(a\right)}+ \frac{2a}{\tanh^3\left(a\right)}\\
	Q_{1}\left(a\right)&=& 6a\csch\left(a\right) -12\coth\left(a\right)\csch\left(a\right) + 12a\csch^3\left(a\right) \\
	J_{1}\left(a\right)&=& -\frac{\sqrt{3}\pi}{64}\left(19-4\cosh\left(a\right) +\cosh\left(2a\right)\right)\sech^4\left(\frac{a}{2}\right)\\
	F_{1}\left(a\right)&=& \frac{3\pi\sqrt{3}}{4}\tanh^2\left(\frac{a}{2}\right)\sech^2\left(\frac{a}{2}\right)\left(\cosh\left(a\right)-\sinh\left(a\right)\right)\\
	K_{1}\left(a\right)&=&  \frac{7}{10} + \frac{3}{2}\csch\left(a\right)\left(a-4\coth\left(a\right)\right) + 6\csch^3\left(a\right)\left(5a-6\coth\left(a\right)\right) + 36a\csch^5\left(a\right) \\
	N_{1}\left(a\right)&=& 9\csch\left(a\right)+  18\csch^3\left(a\right)- 3a\csch\left(a\right)\coth\left(a\right)-18a\csch^3\left(a\right)\coth\left(a\right)\\
	C_{1}\left(a\right)&=& -2\pi\sqrt{3}\csch^3\left(a\right)\sinh^4\left(\frac{a}{2}\right).
	\end{eqnarray}
	\endgroup
Considering the equations of motion associated to the Lagrangian \eqref{9}, we need to impose that $C_{1}\left(a\right)=N_{1}\left(a\right)=0$ in order for the system of differential equations can admit non-trivial solutions. Such convention was made in several works in the literature, as well as considering the asymptotic limit for the function $W_{1}\left(a\to\infty\right)=\frac{3}{4}$ which is the value of characteristic frequency of individual vibrational modes \cite{carlosp,goodnovo,taky,weigel,belova,camp,tadau}.

\begin{figure}[htb!]
		\centering
		\mbox{%
			\subfigure[]
			{
				\includegraphics[scale=0.3]{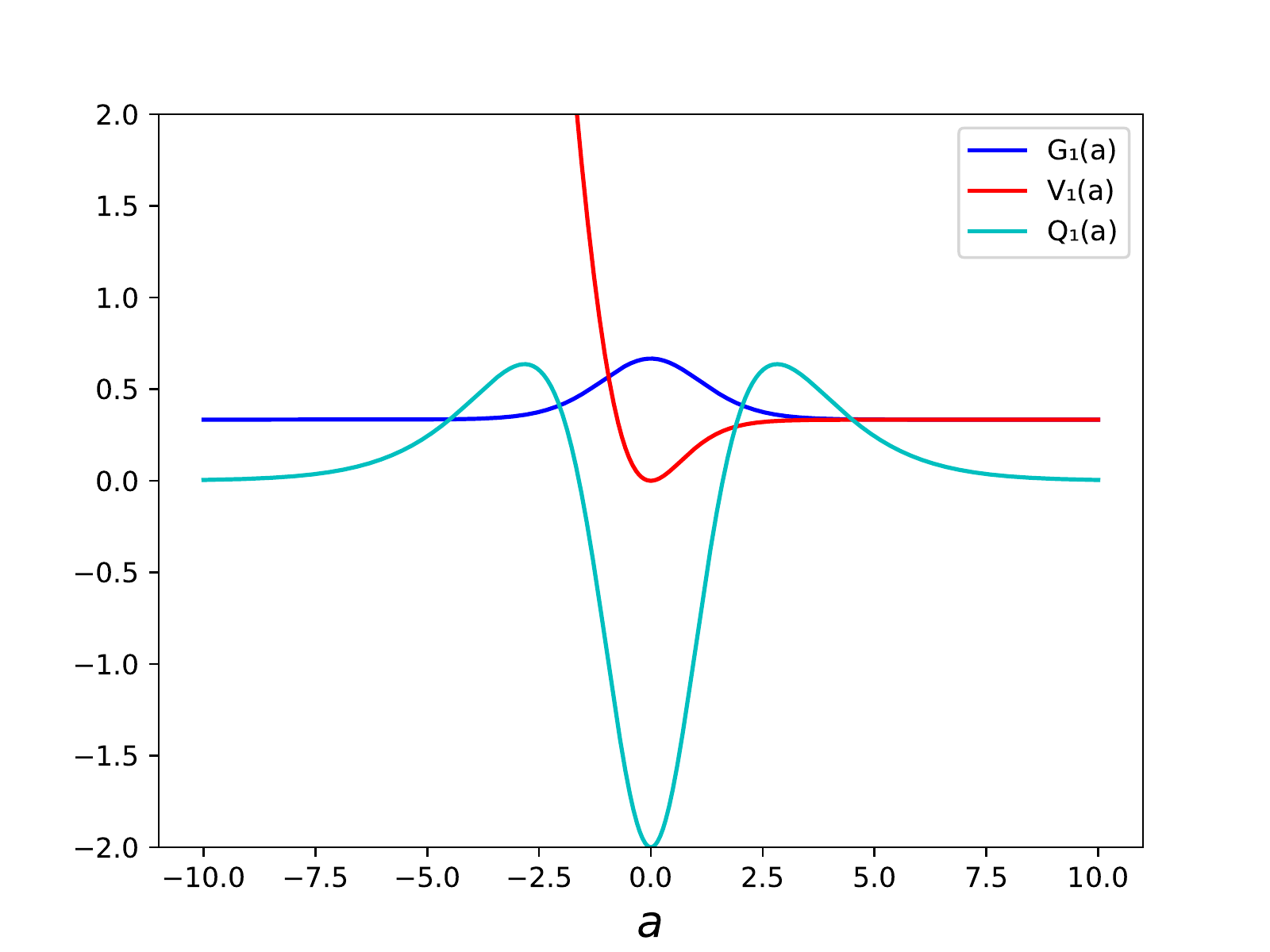}}\qquad
			\subfigure[]
			{
				\includegraphics[scale=0.3]{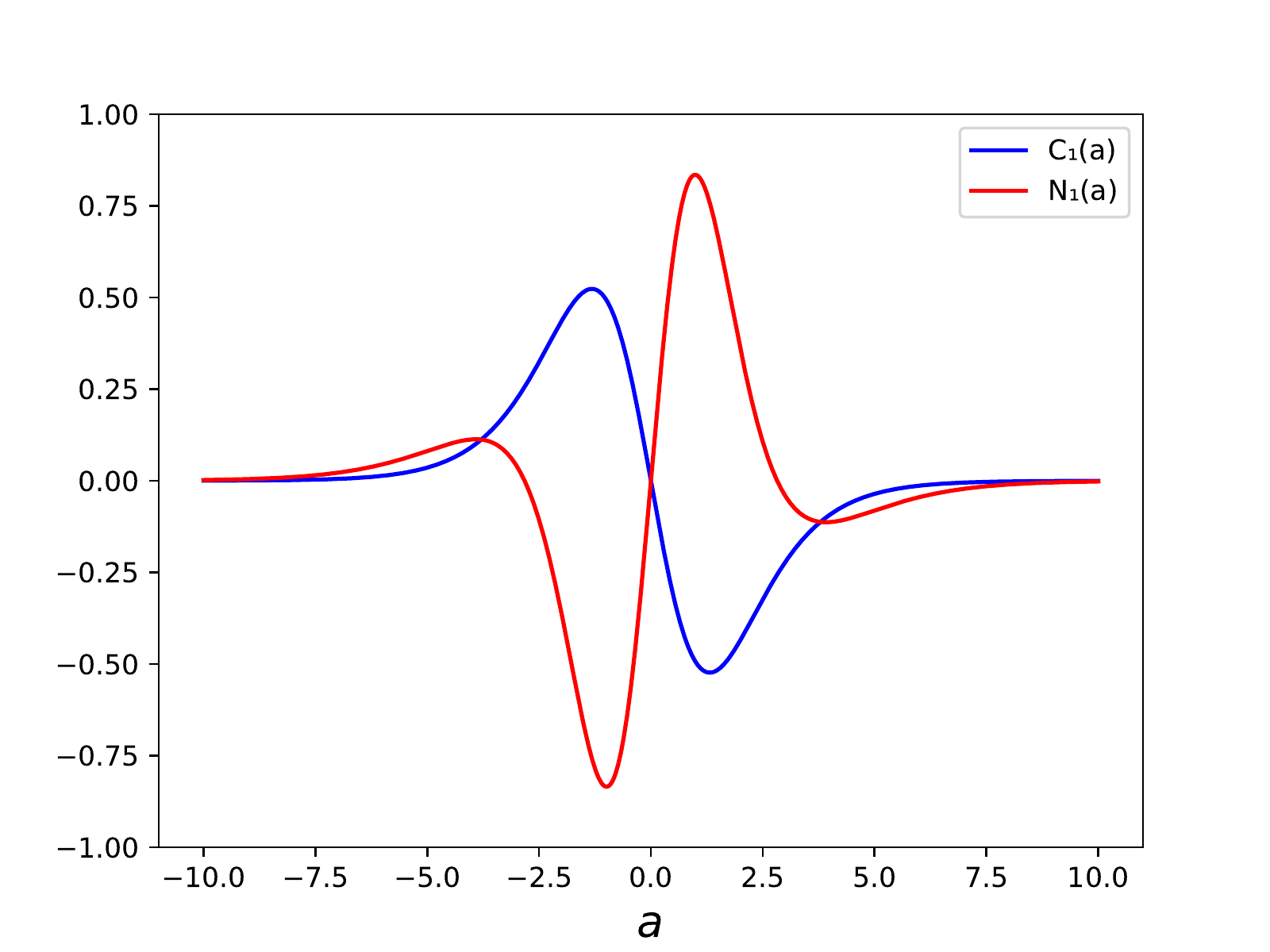}}\qquad 
			\subfigure[] 
			{
				\includegraphics[scale=0.3]{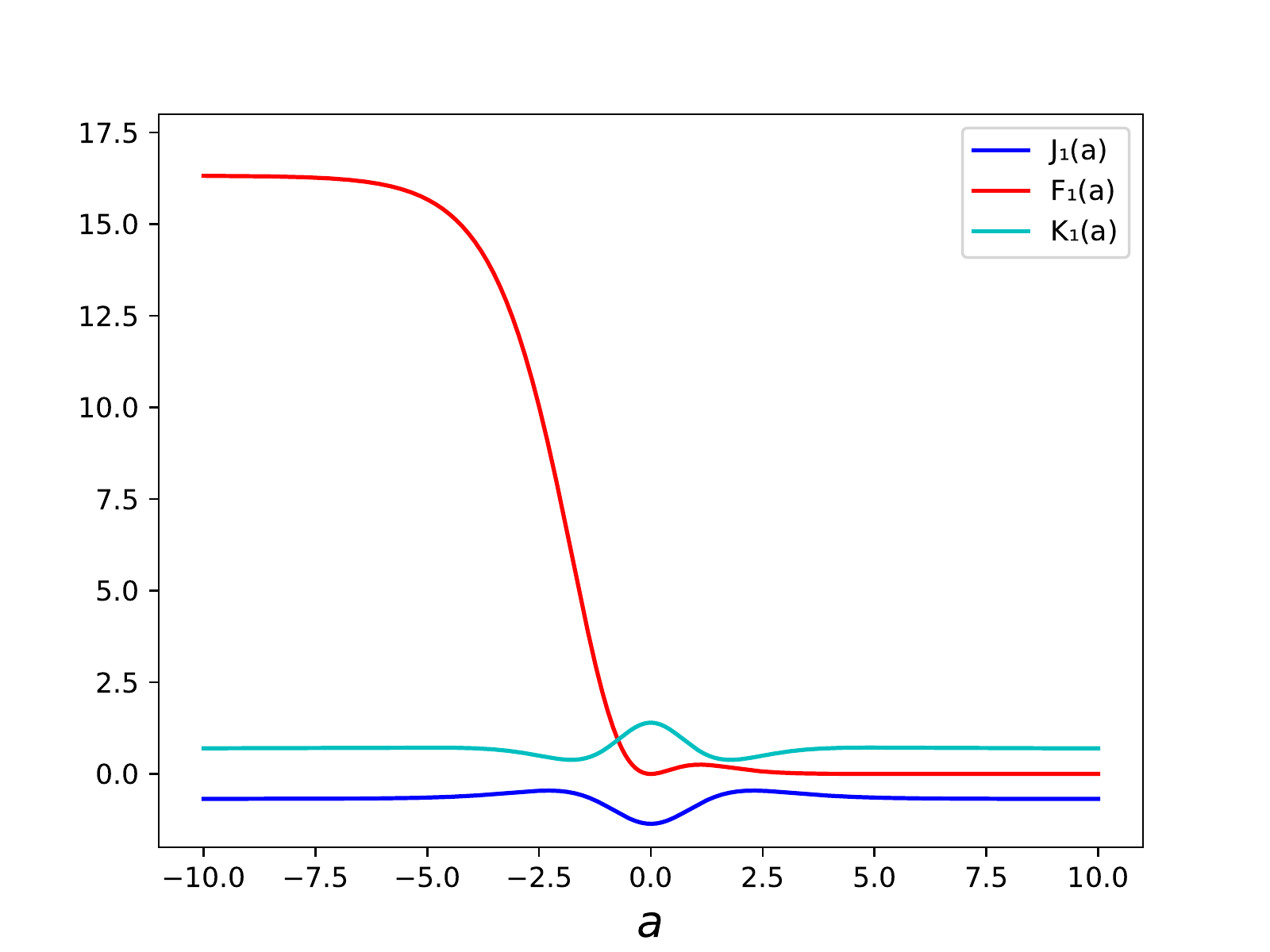}}
		}
		\caption{(a) Behavior of the functions $V_{1}\left(a\right)$, $G_{1}\left(a\right)$ and $Q_{1}\left(a\right)$ where it is straightforward to see that $Q_{1}\left(a\to{0}\right)=-2$ and the divergence of the potential in the limit of $a\to{-\infty}$, (b) the functions $C_ {1}\left(a\right)$ and $N_{1}\left(a\right)$, (c) the functions $J_ {1}\left(a\right)$, $F_ {1}\left(a\right)$ and $K_ {1}\left(a\right)$.}
		\label{fig2}.
	\end{figure}

 Another aspect that has been well explored in recent works is the effort to avoid the \textbf{null-vector} problem, that occurs in the $\phi^4$ polynomial model for instance, and as we can see in the function ${{Q}}_ {1}\left(a\left(t\right)\to{0}\right)=0$ it persists in the present model model\cite{manton1,manton2}. It is worth mentioning that all of the functions that come with the Lagrangean \eqref{8}, except for the potential function $V_{1}\left(a\right)$,  are well behaved, both in their asymptotic limits $a\to\pm\infty$ and in the origin $a\left(t\right)\to{0}$ (\ref{fig2}). By its turn, the potential $V_{1}\left(a\right)$, which is not symmetric with respect to the origin and just as in the case of the model $\phi^4$, diverges when $a\left(t\right)\to{-\infty}$.
	
	For the fullsimulation calculation, we will use as a definition, that is, the average value of the position of the center of mass is written in terms of the energy density. 
	
	\begin{equation}\label{12medio}
	\langle x \rangle = \frac{\int dx\; x\mathcal{E}}{\int dx\; \mathcal{E}},
	\end{equation} where $\mathcal{E}$ is the energy density referring to the configuration \eqref{7}
	\begin{equation}\label{13}
	\mathcal{E} = \frac{1}{2}\left(\frac{\partial \phi}{\partial t}\right)^2+\frac{1}{2}\left(\frac{\partial \phi}{\partial x}\right)^2 + \frac{\phi^2}{2}\left(1-|\phi|\right)^2.
	\end{equation}
\subsection{Analysis of the collective coordinates and fullsimulation for the \textit{anti-kink/kink} scattering}

\begin{figure}[htb!]
		\centering
		\mbox{%
			\subfigure[]
			{\label{akk3}%
				\includegraphics[scale=0.3]{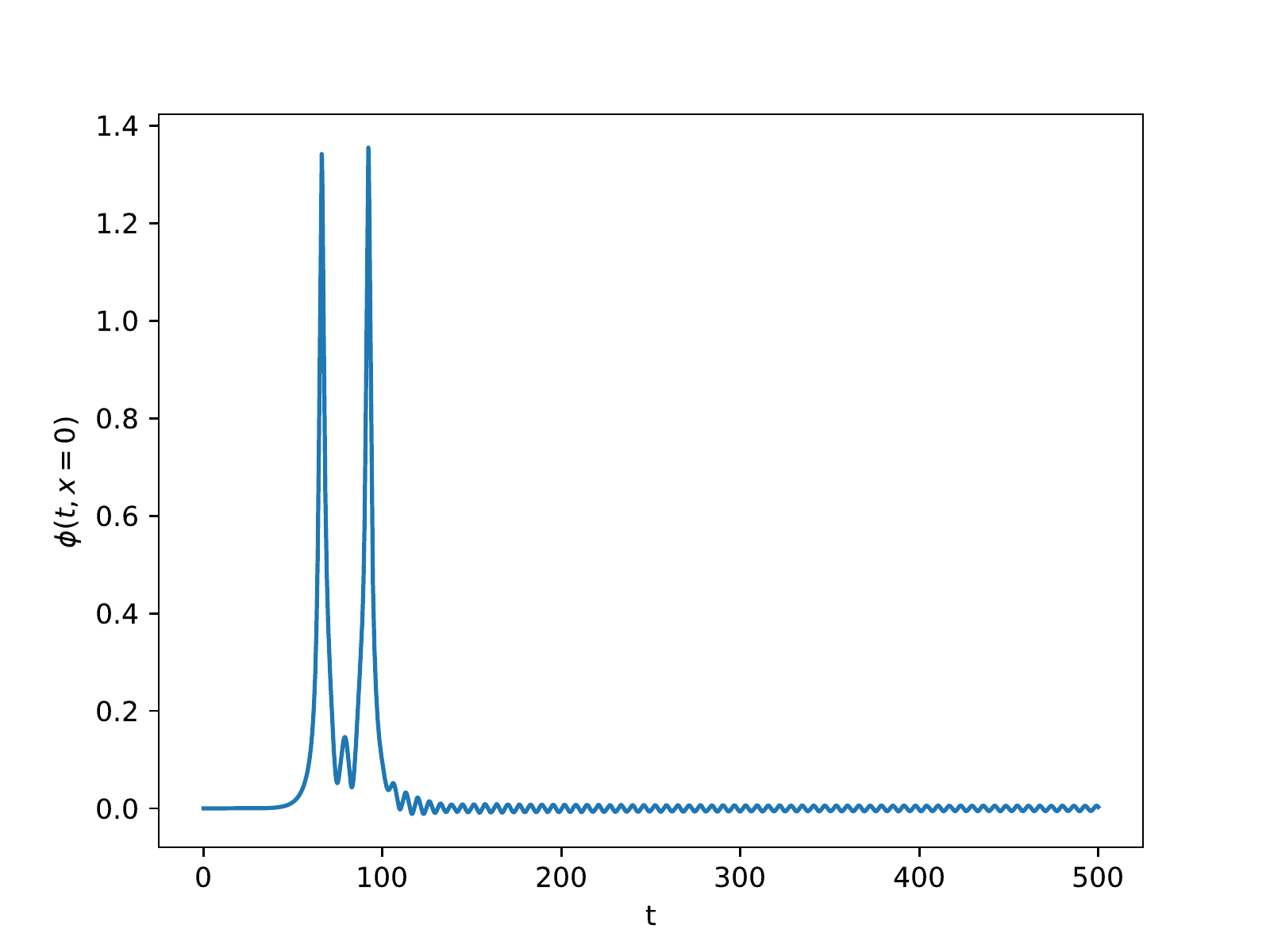}}\qquad
			\subfigure[]
			{\label{akk2}%
				\includegraphics[scale=0.3]{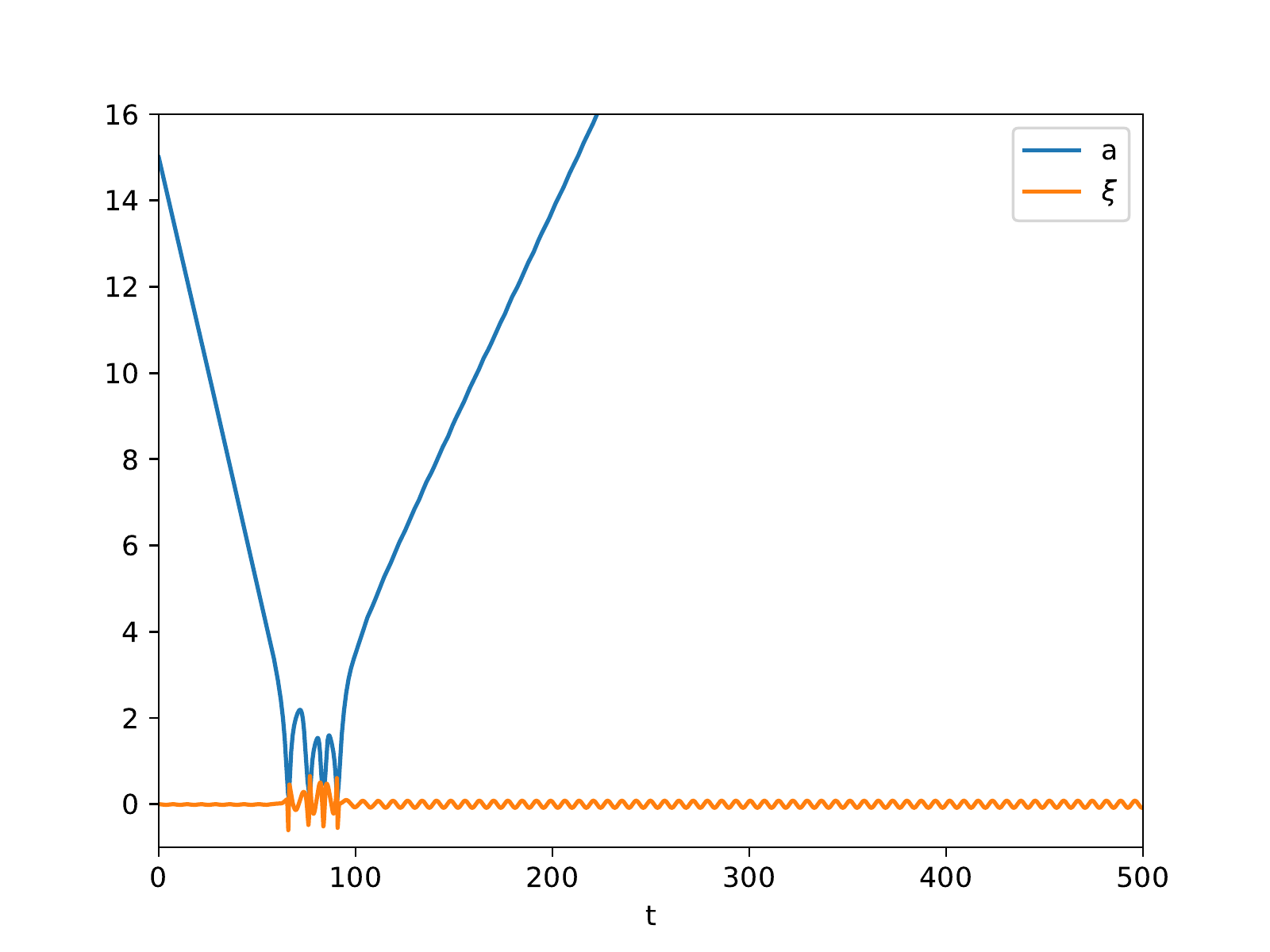}}\qquad
			\subfigure[]
			{\label{akk1}%
				\includegraphics[scale=0.3]{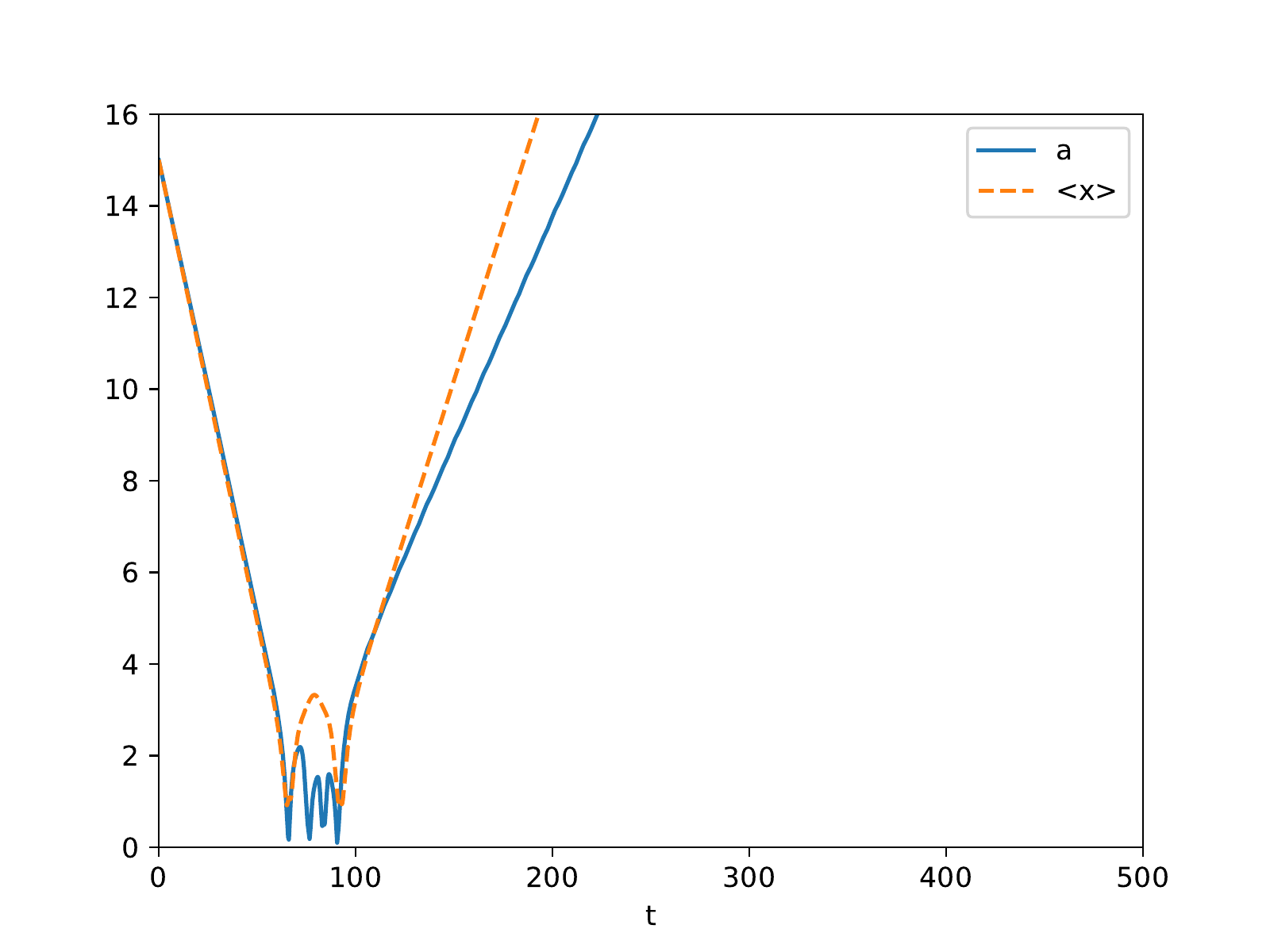}}
		}
		\caption{(a) Time evolution for the \textit{anti-kink/kink} scattering with an initial speed of $v_{ini}=0.201$, (b) energy transfer between the internal modes, being $a\left(t\right)$ translation and $\xi\left(t\right)$ vibration and (c) comparison between fullsimalation and collective coordinates for same initial velocity $v_{ini}=0.201$.}
		\label{figakkcc}.
	\end{figure}

For the configuration of \textit{anti-kink/kink}, we perform the dynamics for fullsimulation and verify that for scatterings with velocities above the critical velocity, $v_{ini}>v_{crit}=0.2599$, we only have inelastic collisions. That is, the \textit{anti-kink/kink} perform only the first contact and leave for asymptotics. On the other hand, for scatterings with an initial velocity below $v_{ini}<v_{crit}=0.2599$, we have the formation of so-called bion states, which remain emitting radiation for a long period of time until they annihilate. Thus, we can also verify the characterization of resonance states in which it is possible to observe the formation of a fixed number of resonance windows that can vary from two to four \ref{akk3}.

Regarding the dynamics of collective coordinates, as we well know from previous works \cite{camp,belova,good,moshi,samuel,tadau} this technique is an approximative method, and therefore, only for some scattering velocities do the results agree reasonably with the fullsimulation . In this work, it is no different, notice in \ref{akk1} that we did the scattering for an initial velocity $v_{ini}=0.201$ and the result of the dynamics in a qualitative way agrees with the fullsimulation. Also, in figure \ref{akk1}, it is observed that the number of jumps represented in the collective coordinates should not be seen as resonance windows. Finally, we have the energy transfer from the internal translation mode to the internal vibration mode \ref{akk2}.
	
\section{\textbf{kink/anti-kink} scattering}
	
	In the present section, we will study the scattering of the \textit{kink/anti-kink} configuration. We will take as a starting point the field configuration below
	
	\begin{eqnarray}\label{14}
	\phi= \phi_{K}^{\left(0,1\right)}\left(x + a(t)\right) + \phi_{\bar{K}}^{\left(1,0\right)}\left(x - a(t)\right) -1 + \xi\left(t\right)\left[\chi_{K}^{\left(0,1\right)}\left(x+a(t)\right) - \chi_{\bar{K}}^{\left(1,0\right)}\left(x-a(t)\right)\right],
	\end{eqnarray} with the vacuum solutions $\phi_{K}^{\left(0.1\right)}$, $\phi_{\bar{K}}^{\left(1.0\right)}$ and their respective normal vibration modes: $\chi_{K}^{\left(0.1\right)}$ and $\chi_{\bar{K}}^{\left(1.0\right)}$. As in the previous case, for simplicity, we will only consider terms of up to second order in the amplitude of the vibrational modem $O\left(\xi^2\right)$.
	
	The effective Lagrangian for the scattering is given by
	
	\begin{eqnarray}\label{15}
	L_{2}\left(a,\dot{a},\xi,\dot{\xi}\right)&=&\frac{1}{2}\left[\bar{M}+I_{2}\left(a\right)+\xi{J_{2}\left(a\right)}+\xi^2{K_{2}\left(a\right)}\right]\dot{a}^2 +\frac{1}{2}\left[{2+Q_{2}\left(a\right)}\right]\dot{\xi}^2 + \\\nonumber
	&+& \left(C_{2}\left(a\right)+\xi{N_{2}\left(a\right)}\right)\dot{a}\dot{\xi} -\left(V_{2}\left(a\right)-\xi{F_{2}\left(a\right)}+\xi^2{W_{2}\left(a\right)}\right).
	\end{eqnarray} The rest mass for \textit{kink} and \textit{anti-kink} configurations are the same. The auto-frequencies are also identical in asymptotics and therefore
	$W_{1}\left(a\to \infty\right)=W_{2}\left(a\to \infty\right)=\omega_{Hybrid}^2=\frac{3}{4}$ .
	
	In this way, some functions for, both \textit{anti-kink/kink} and \textit{kink/anti-kink} scattering, are equal $I_{1}\left(a\right )= I_{2}\left(a\right)$, $V_{1}\left(a\right)= V_{2}\left(a\right)$, $Q_{1}\left(a \right)= Q_{2}\left(a\right)$, $K_{1}\left(a\right)= K_{2}\left(a\right)$ and $N_{1}\left (a\right)= N_{2}\left(a\right)$. By its turn, the other functions are anti-symmetric in relation to the scattering configuration change: $J_{1}\left(a\right)= -J_{2}\left(a\right)$, $F_{1}\left(a\right)= -F_{2}\left(a\right)$ e $C_{1}\left(a\right)= -C_{2}\left(a\right)$. 
	
	An important observation is that for both scattering cases, the \textbf{null-vector} problem is maintained, and, in some way, the existence of this term in the Lagrangian of the collective coordinates is relevant for the dynamics to be compatible with fullsimulation.
	
This analysis goes against how occurs the combination between the internal vibrational modes constructed in equations \eqref{7},\eqref{14}.  Notice that by construction the ansatz $\left[\chi_{\bar{K}}^{\left(1.0\right)}\left(x+a(t)\right)-\chi_{K}^ {\left(0,1\right)}\left(x-a(t)\right)\right]$, both in the limit of $a\to{0}$ and in the limit of $a\left(t\right )\to{\infty}$, the terms cancel each other out. Therefore, at first, this choice is inconsistent if we consider that, at the limit $a\left(t\right)\to{0}$, the energy exchange between the internal modes must be a maximum, on the other hand, it is this ansatz that shows consistent with fullsimulation and predicts the resonance windows.
	
	On the other hand, if we consider a sum of the vibrational modes  in the ansatz, $\left[\chi_{\bar{K}}^{\left(1,0\right)}\left(x+a(t)\right)+\chi_{K}^{\left(0,1\right)}\left(x-a(t)\right)\right]$, we will see that in the limit $a\left(t\right)\to\infty$, the contribution of these modes is null, as expected since the objects are free. On the other hand, at the limit of $a\left(t\right)\to{0}$ the contribution of these vibrational modes is maximum, as we expect. However, when these objects are scattered, there is no resonance window, thus being incompatible with fullsimulation.

\subsection{Description of the collective coordinates and fullsimulation for the \textit{kink/anti-kink} scattering}

\begin{figure}[htb!]
		\centering
		\mbox{%
			\subfigure[]
			{\label{kak1}%
				\includegraphics[scale=0.3]{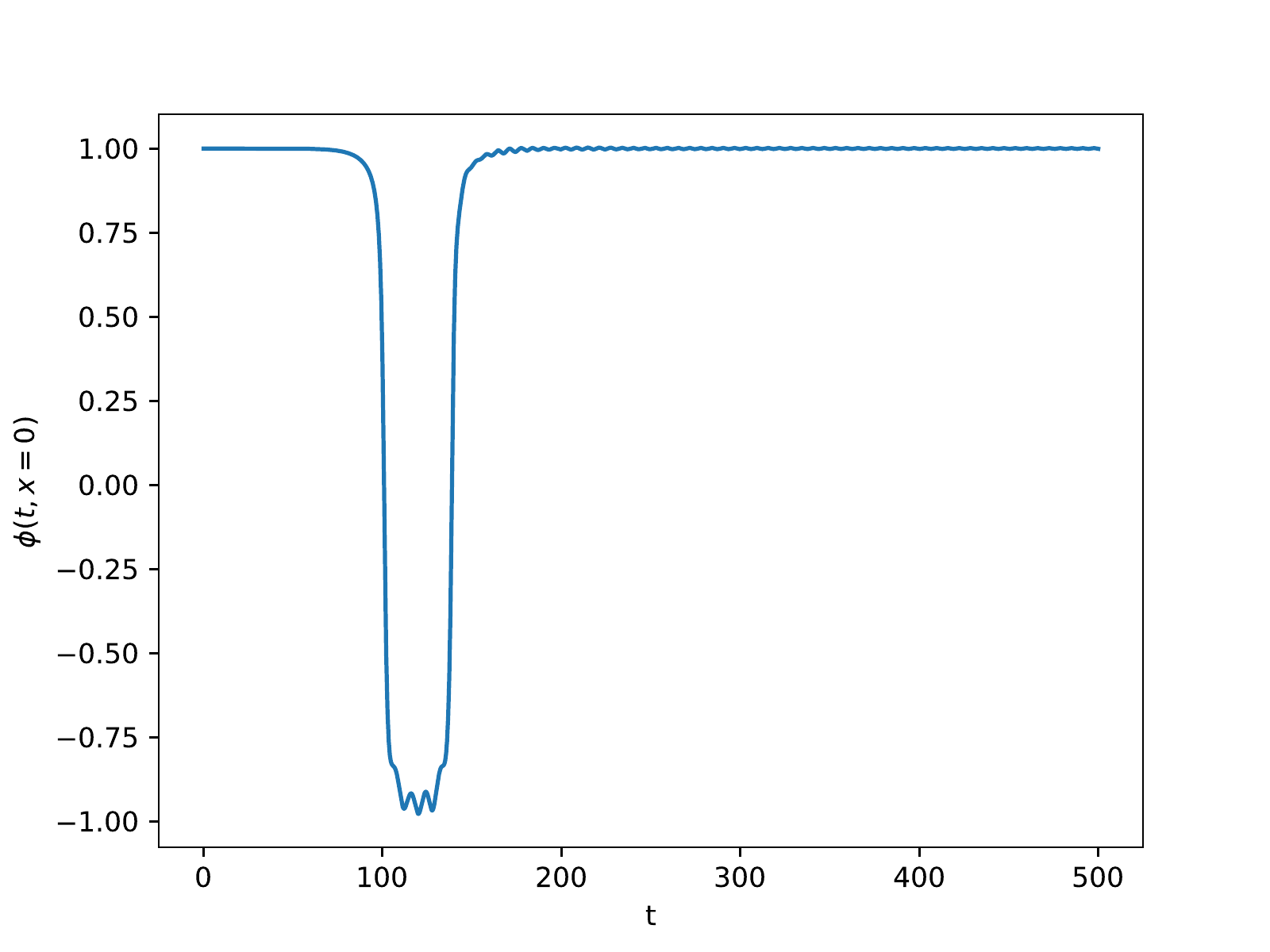}}\qquad
			\subfigure[]
			{\label{kak2}%
				\includegraphics[scale=0.3]{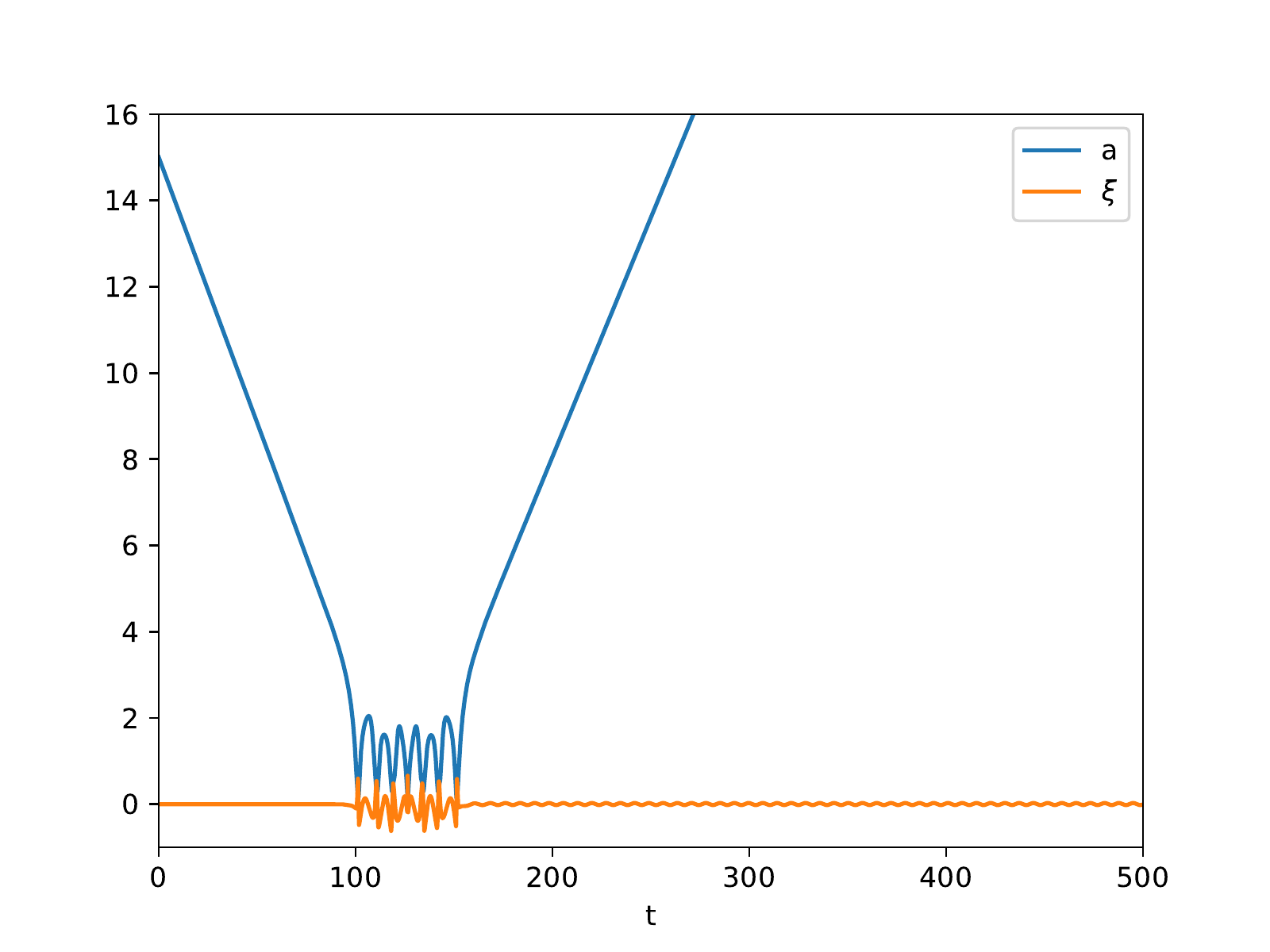}}\qquad
			\subfigure[]
			{\label{kak3}%
				\includegraphics[scale=0.3]{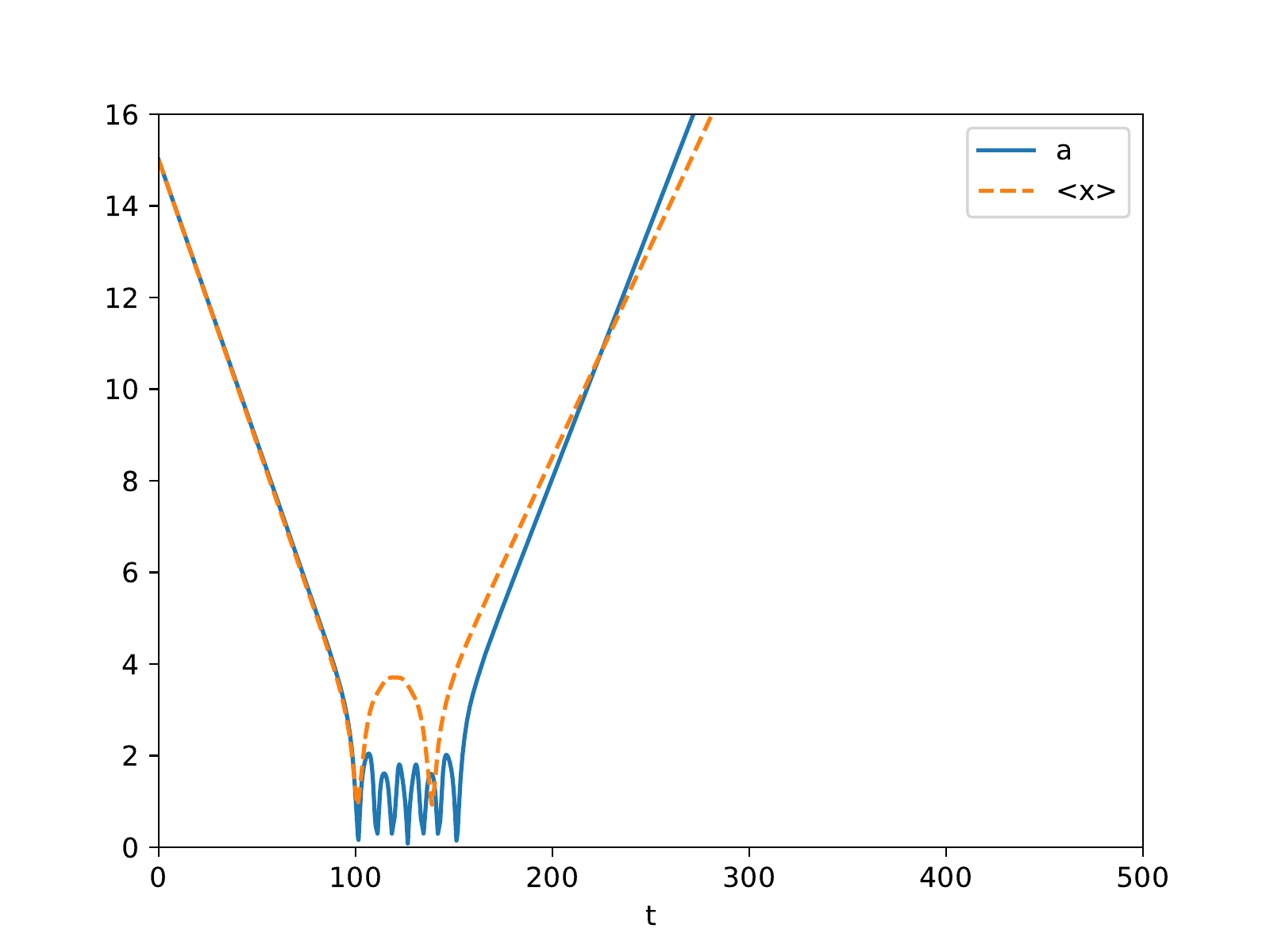}}
		}
		\caption{(a) Time evolution for the \textit{kink/anti-kink} scattering with an initial speed of $v_{ini}=0.124$, (b) energy transfer between the internal modes, being $a\left(t\right)$ translation and $\xi\left(t\right)$ vibration and (c) comparison between fullsimalation and collective coordinates for same initial velocity $v_{ini}=0.124$.}
\label{figkakcc}.
\end{figure}

The scattering was performed for the \textit{kink/anti-kink} configuration in which we verified the same critical velocity $v_{crit}= 0.152$ obtained in D. Bazeia et al \cite{hibrid}. For scatterings with an initial velocity greater than the critical velocity $v_{ini}>v_{crit}=0.152$, we have the inelastic collisions. On the other hand, the formation of bion states is verified for many velocities below the critical velocity.

There are no indications of the formation of resonance windows in the present scattering configuration, but states with only one jump. However, before the \textit{kinks} go the asymptotics, a certain number of oscillations are found. In figure \ref{kak1}, we see the formation of such state for the scattering with an initial velocity of $v_{ini}=0.124$. That is, we have the formation of three oscillations before the \textit{kinks} go the asymptotics. We also performed the collective coordinates method for this initial velocity, and it is in agreement with the fullsimulation \ref{kak3}. Furthermore, we can see in the figure \ref{kak2} the energy redistribution between the internal modes.

\section{Analysis of the null-vector problem}
	
	Recently, the authors in the works \cite{manton1,manton2} proposed the introduction of an attenuator function $f\left(a\right)$ in the ansatz of the collective coordinates for the potential $\phi^4$ \cite{weigel,taky, carlosp} in order to solve the \textbf{null-vector} problem. An interesting aspect is that, it appears the idea that when the \textbf{null-vector} problem is corrected, a singularity is necessarily created in the moduli-space. This fact was also verified for the model of the present study, in which we performed the introduction of the function $f\left(a\right)=\tanh\left(a\right)$ both in the ansatz of the scattering of \textit{anti-kink /kink} \eqref{pv1} as for the scattering \textit{kink/anti-kink} \eqref{pv2}

\begin{eqnarray}\label{pv1}
\phi= \phi_{\bar{K}}^{\left(1,0\right)}\left(x+ a(t)\right)+\phi_{K}^{\left(0,1\right)}\left(x -a(t)\right) + \frac{\xi\left(t\right)}{f\left(a\right)}\left[\chi_{\bar{K}}^{\left(1,0\right)}\left(x+a(t)\right)-\chi_{K}^{\left(0,1\right)}\left(x-a(t)\right)\right],
\end{eqnarray}

\begin{eqnarray}\label{pv2}
\phi= \phi_{K}^{\left(0,1\right)}\left(x + a(t)\right) + \phi_{\bar{K}}^{\left(1,0\right)}\left(x - a(t)\right) -1 + \frac{\xi\left(t\right)}{f\left(a\right)}\left[\chi_{K}^{\left(0,1\right)}\left(x+a(t)\right) - \chi_{\bar{K}}^{\left(1,0\right)}\left(x-a(t)\right)\right].
	\end{eqnarray}

\begin{figure}[htb!]
		\centering
		\mbox{\subfigure[]
			{\label{akk02}
			\includegraphics[scale=0.47]{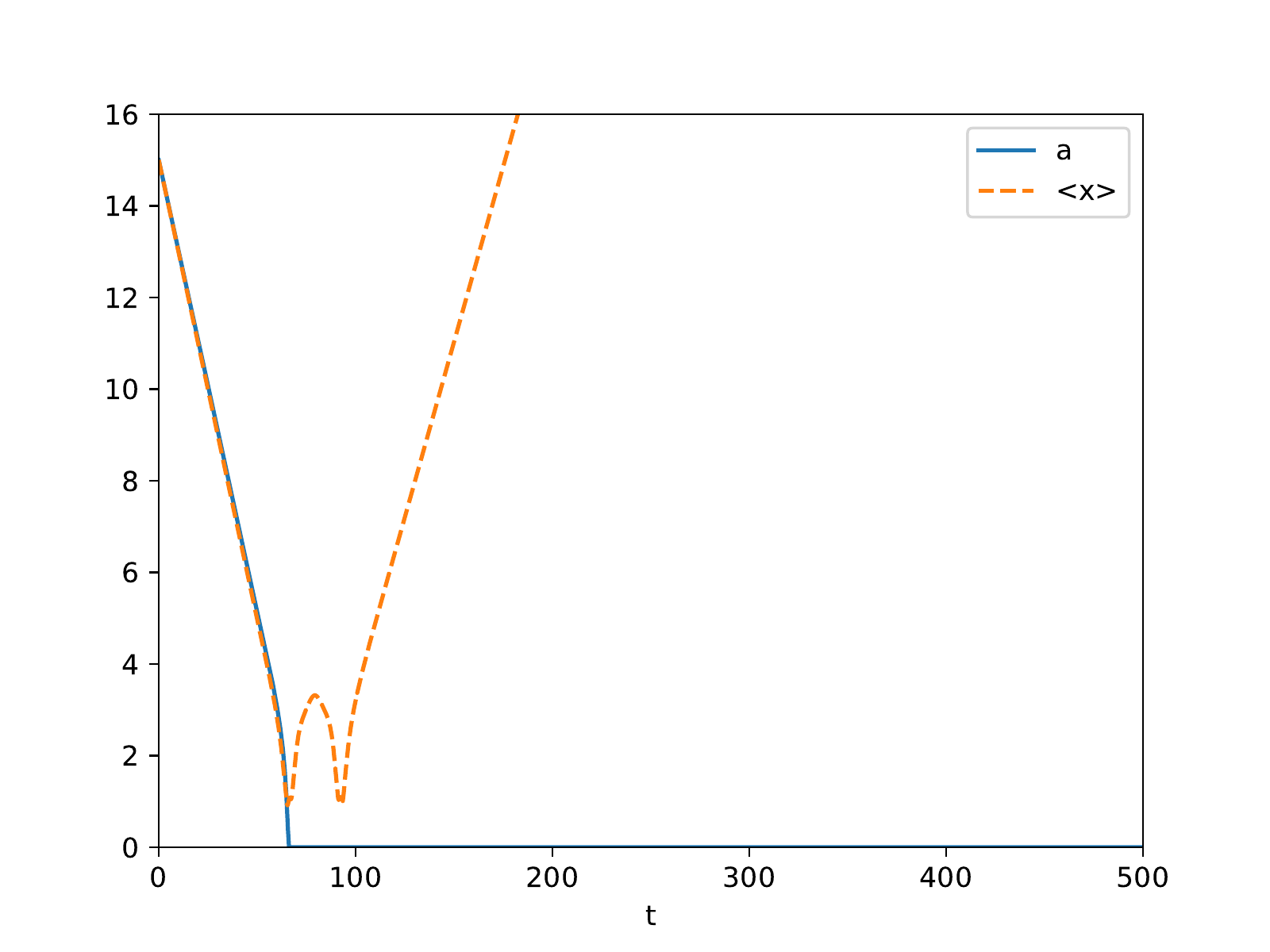}}\qquad
			\subfigure[]
			{\label{field02}
			\includegraphics[scale=0.47]{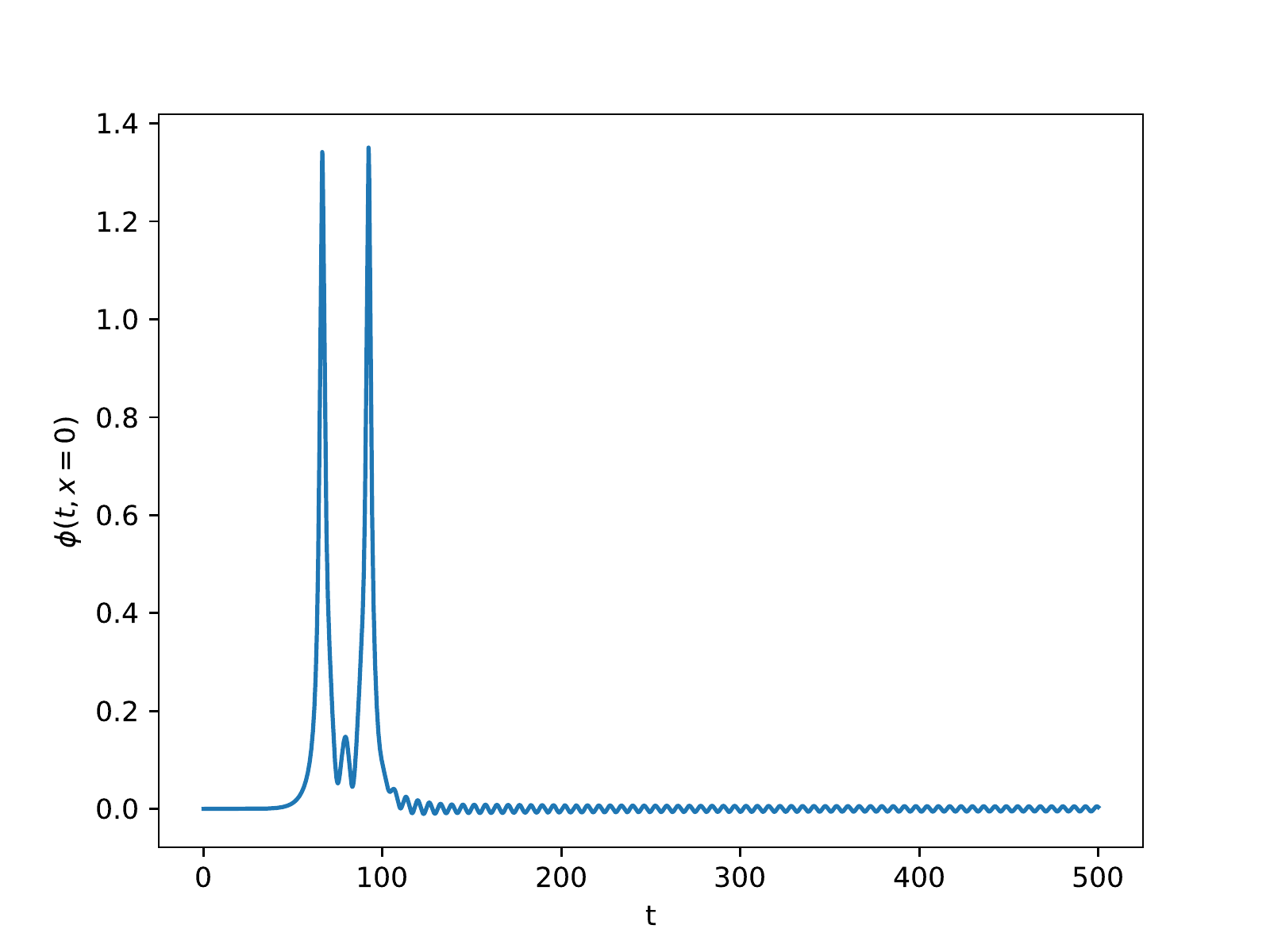}}
}
\caption{(a) \textit{anti-kink/kink} scattering with an initial velocity of $v_{ini}=0.2$ (b) Time evolution of the solution at the origin, with the formation of two resonance windows.}
\label{fignmakk}
\end{figure}

The initial configuration of the field for the \textit{anti-kink/kink} scattering, given in equation \eqref{pv1}, is represented in the figure (\ref{fignmakk}). In the figure on the right \ref{field02}, we have the temporal evolution of the field in which we can see the formation of two resonance windows. While in the left figure \ref{akk02}, we have the comparison between the fullsimulation and the collective coordinates. Notice also that,  after the occurrence of the first collision, the dynamics associated with the collective coordinate of translation is frozen in time, just as it happens for the case $\phi^4$ polynomial. Such effect occurs due to the manifestation of the singularity in the moduli-space predicted in N. Manton et al. \cite{manton1}.

\begin{figure}[htb!]
		\centering
		\mbox{\subfigure[]
			{\label{kak0124}
			\includegraphics[scale=0.47]{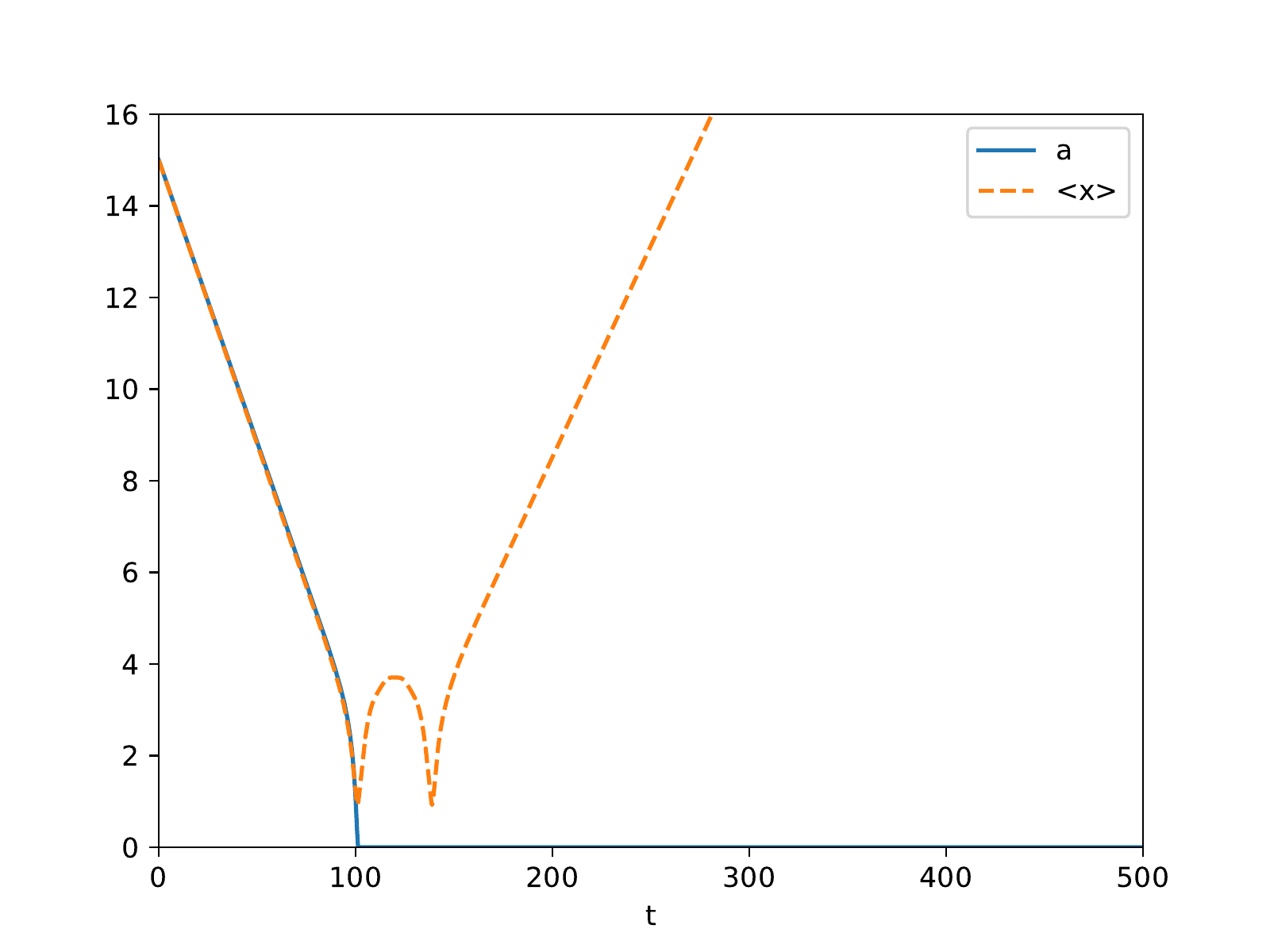}}\qquad
			\subfigure[]
			{\label{camp0124}
			\includegraphics[scale=0.47]{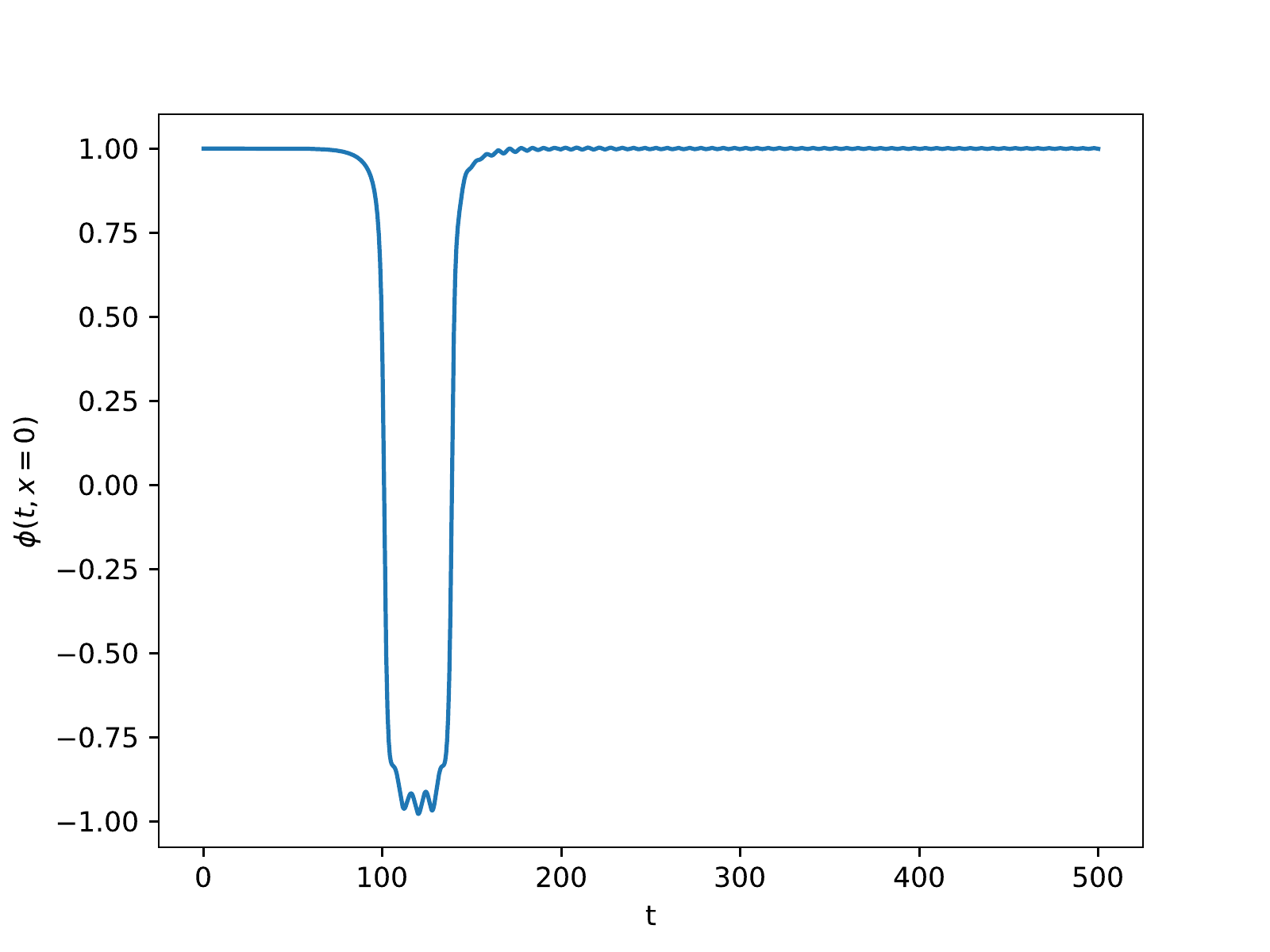}}
}
\caption{(a) \textit{kink/ant-kink} scattering with an initial velocity of $v_{ini}=0.124$ (b) Time evolution of the solution at the origin, with the formation of three oscillations before the separation.}
\label{fignmkak}
\end{figure}

Analogously, we perform the \textit{kink/anti-kink} scattering for the field configuration given by equation \eqref{pv2} and we can observe the temporal evolution for the initial velocity of $v_{ini}=0.124$ and the formation of three oscillations before the kinks come out to the asymptotic \ref{camp0124}. For the figuration \ref{kak0124} once again, we have the solution freezing as an effect of the singularity.

\section{Conclusion}

We carried out the study of collective coordinates referring to the hybrid model for both the \textit{anti-kink/kink} scattering and the \textit{kink/anti-kink} scattering. We saw that it has a connection with the potentials $\phi^4$ and $\phi^6$ and that just as happens in the case of the potential $\phi^6$, it has two topological sectors. We took advantage of the mapping with the $\phi^4$ polynomial model to make use of the integration technique used in the work \cite{carlosp} and build the Lagrangian functional only in terms of the dynamic parameters and their derivatives.

Since both the Lagrangian functionals for the \textit{anti-kink/kink} scattering and the \textit{kink/anti-kink} scattering have remained with the \textbf{null-vector} problem, we investigate the resolution of this problem by following the redefinition of the ansatz presented by N. Manton et al. \cite{manton2}, and we conclude that the moduli space becomes singular, just like the polynomial case \cite{manton1}.

As a future perspective, it would be interesting to investigate whether, for any well-behaved attenuator function in asymptotic limits that can be introduced in the correction of the \textbf{null-vector} problem, necessarily produce a singularity in the moduli-space\cite{manton1}. Because as the initial interest was to correct the \textbf{null-vector} problem and thus verify how the relocation of energy between the internal modes occurs, the creation of a singularity does not contribute to elucidating the problem since the solution of the collective coordinates is frozen in time.

\begin{acknowledgments}
This work made use of the Virgo Cluster at Cosmo-ufes/UFES, which is funded by FAPES (Fundação de Amparo à Pesquisa e Inovação do Espírito Santo) and administered by Renan Alves de Oliveira. C.F.S. Pereira thanks the financial support provided by the Coordination for the Improvement of Higher Education Personnel (CAPES).  E.S.C. Filho thanks the supported by the Center for Research and Development in Mathematics and Applications (CIDMA) through the Portuguese Foundation for Science and Technology (FCT - Fundação para a Ciência e a Tecnologia), references UIDB/04106/2020 and UIDP/04106/2020. T. Tassis thanks the financial support provided by the Federal University of ABC (UFABC).
\end{acknowledgments}

\nocite{*}
\bibliography{electrovac}

\end{document}